\documentclass[epj]{svjour}
\usepackage{graphics}
\usepackage[dvips]{color}
\usepackage{epsfig}

\newcommand{\ds}{\displaystyle}
\newcommand{\dsf}{\ds\frac}
\newcommand{\Tr}{\mbox{Tr}}
\newcommand{\re}[1]{(\ref{#1})}
\newcommand{\no}{\nonumber}
\newcommand{\intrT}{\int\limits_0^\infty dr\,r^{2}
\int\limits_0^\pi\!S_\theta\,d\theta\,}
\begin{document}
\title{Neutron-proton mass difference in finite nuclei
and the Nolen-Schiffer anomaly}

\author{
 Ulf-G. Mei{\ss}ner\inst{1,2}%\thanks{e-mail: meissner@itkp.uni-bonn.de}
   \and 
 A.~M.~Rakhimov\inst{3,4}
    \and 
 A.~Wirzba\inst{2}%\thanks{e-mail: a.wirzba@fz-juelich.de}
    \and 
 U.~T.~Yakhshiev\inst{2,5}%\thanks{e-mail: u.yakhshiev@nuuz.uzsci.net}
}                     % Do not remove
\institute{
Helmholtz-Institut f{\" u}r Strahlen- und Kernphysik (Theorie),
D-53115, Universit{\" a}t Bonn, Germany
\and
Forschungszentrum
J{\" u}lich, Institut f{\" u}r Kernphysik
(Theorie),  D-52425  J{\" u}lich, Germany
\and
Institute of Nuclear Physics, Academy of Sciences of
Uzbekistan, Tashkent-132, Uzbekistan
\and
 Institute of Physics and Applied Physics,
Yonsei University, Seoul, 120-749, Korea
\and
Physics Department and Institute of Applied Physics,
National University of Uzbekistan, Tashkent-174, Uzbekistan}
\authorrunning{U.-G. Mei{\ss}ner et al.}
\titlerunning{Neutron-proton mass difference in finite nuclei ...}
\date{Received: date / Revised version: date}
\abstract{ The neutron-proton mass difference in finite nuclei is
  studied in the framework of a medium-modified Skyrme model.  
The possible interplay between the effective nucleon mass in finite nuclei and 
the Nolen-Schiffer anomaly is discussed. 
In particular, we find that a correct description of the properties of 
mirror nuclei  leads to a stringent restriction of  possible
modifications of the nucleon's effective mass in nuclei.
}

 % end of PACS codes

%\keywords{Skyrmions, protons and neutrons, nuclear matter.}

%\today
%
\maketitle
%

%%%%%%%%%%%%%%%%%%%%%%%%%%%%%%%%%%%%%%%%%%%%%%%%%%%%%%%%%%%%%%%%%%%%%%%%%%%%%%
\section{Introduction}
\label{sect:intro}
\setcounter{equation}{0}

An essential quantity in nuclear physics, the effective
neu\-tron-proton mass difference in nuclear matter, 
$\Delta m_{\rm  np}^*$, is still not known
empirically~\cite{Lunney03}\footnote{Thereafter an asterix in an
  expression indicates that the corresponding quantity is explicitly 
  medium-modified.}.
On the other hand,  there exist very different theoretical predictions
of this quantity 
for isospin-asymmetric nuclear matter 
\cite{Bo91,Zu01,Zu05,Cha97,Le06,LQ88,vD05,vD06,Ku97,Liu1,Gr00,Hof0,Ts99,Ch07} 
where the results differ both qualitatively and quantitatively. 
Such studies of the effective neutron-proton mass difference inside 
nuclei may be relevant to resolve the Nolen-Schiffer anomaly (NSA) in
nuclear physics~\cite{Nolen:1969ms,Shlomo} and for
applications in 
nuclear astrophysics~\cite{Li:1997px,Baran:2004ih,Steiner:2004fi}. 
Although there
are many theoretical approaches devoted to the explanation of the
NSA~\cite{Sh82,Agr01,He89,Ha90,Wi85,Co90,Sh94xz,Horo,Suz,Sch93xt,Dr94,Ad91}
this phenomenon is still not fully understood.

In this context, we have recently proposed an 
effective Lagrangian~\cite{Meissner:2007ya} which can be applied to 
investigate isospin-breaking effects in the baryonic sector of the 
Skyr\-me model under medium modifications. 
In ref.~\cite{Meissner:2007ya} the single nucleon properties and 
 $\Delta m_{\rm np}^*$ are studied in infinite, asymmetric 
nuclear matter with a spatially constant density, where
the surrounding medium influences were taken into
account as  external parameters.
Moreover, the model can be extended to  
the evaluation of nucleon properties 
in {\em finite} nuclei. Such kind of studies, 
however with only partial 
isospin-splitting effects, 
have already been performed in 
refs.~\cite{Yakhshiev:2001ht,Yakhshiev:2002sr}.
It will be interesting to extend these studies to the
strong and electromagnetic isospin-breaking effects in 
finite nuclei. 
Consequently, in the present work we 
consider the neutron-proton mass difference 
in finite nuclei by studying the single-nucleon 
effective properties and by trying 
to explain the Nolen-Schiffer anomaly within a
medium-modified Skyrme model.

The paper is organized as follows: In sect.~\ref{sect:model} we review 
the model Lagrangian which was 
adapted in ref.~\cite{Meissner:2007ya} to skyrmions in 
asymmetric nuclear matter.
Furthermore, we 
specify  the input parameters and
the medium functionals, which express the influence of the surrounding 
environment onto the single-nucleon properties, together with
the parameterization of the neutron- and proton-density distributions.
Section~\ref{sect:parI} centers on the
spherically symmetric case where
a classical skyrmion  is located at the center of a spherical  nucleus.  
In particular, we contrast the exact solution of the equations
of motion with  a variational approximation and compare in
sect.~\ref{sect:res_sym} the results of both 
calculations. Next, in
sect.~\ref{sect:nonsph_sol} we extend our studies to the
nonspherical case where the  nucleons are located at some finite 
distance
from the center of nucleus. We discuss the consequences of the
nonspherical  case for the time-dependent Lagrangian, the
appropriate classical solutions and the quantization procedure.
In sect.~\ref{sect:nonsph-EMff}, 
we present the pertinent calculations for the
electromagnetic part of the neutron-proton mass difference in finite nuclei
and in sect.~\ref{sect:res_nonsym} we
discuss the results of the
nonspherical approximation. 
Our discussions of the Nolen-Schiffer anomaly 
are presented in sect.~\ref{sect:NSA} and a  summary of the
results is given in sect.~\ref{sect:sum}.  Finally and for
completeness, two appendixes are added which elaborate on the nonspherical
scenario: the first is devoted to the
mass functional and moments of inertia 
(appendix~\ref{app:mass_momin}) and the second to the pertinent
charges  (appendix~\ref{app:Charges}).

\section{Lagrangian of the model}
\label{sect:model}

We start with the Skyrme-model Lagrangian presented 
in ref.~\cite{Meissner:2007ya}
\begin{eqnarray}
{\cal L}^*\!&=&\! {\cal L}^*_{\rm sym}+{\cal L}^*_{\rm as}\,,
\label{lmedbeg}\\
{\cal L}^*_{\rm sym}\!&=&\!{\cal L}_2^*+{\cal L}_4
                        +{\cal L}_{\chi{\rm SB}}^*\,,\\
{\cal L}^*_{\rm as} &=&
\Delta{\cal L}_{\rm mes}+\Delta{\cal L}_{\rm env}^*\,,
\label{asym}\\
{\cal L}_2^*\!&=&\!\dsf{F_\pi^2}{16}
\Big\{
\left(1+\dsf{\chi_{s}^{02}}{m_\pi^2}\right)
\Tr\left(\partial_0U\partial_0U^\dagger\right)\no\\
&&\mbox{}-\left(1-\chi_{p}^0\right)
\Tr(\vec{\nabla} U\cdot\vec{\nabla} U^\dagger)
 \Big\},
\label{L2}
\\
{\cal L}_4&=&\dsf{1}{32e^2}\,\Tr\,[U^\dagger\partial_\mu U,
U^\dagger \partial_\nu U]^2\,,\\
{\cal L}_{\chi {\rm SB}}^*&=&-\dsf{F_\pi^2 m_{\pi}^2}{16}
\big(1+m_\pi^{-2}{\chi_{s}^{00}}\big)
\no\\&&\mbox{}\times 
\Tr\left[(U-1)(U^\dagger-1)\right]\,,
\label{L_chsb}\\
\Delta{\cal L}_{\rm mes}\!&=&\!-\dsf{F_\pi^2}{16}\sum_{a=1}^2{\cal M}_-^2
\Tr(\tau_aU)\Tr(\tau_aU^\dagger),\quad\,\,\\
\Delta{\cal L}_{\rm env}^*\!&=&\!-\dsf{F_\pi^2}{16}\sum_{a,b=1}^2
{\varepsilon_{ab3}
\dsf{\Delta\chi_{s}+\Delta\chi_{p}}{2m_\pi}
}\no\\
&&\qquad\mbox{}\times \Tr(\tau_a U)\Tr(\tau_b\partial_0 U^\dagger)\,,
\label{lmedend}
\end{eqnarray}
where Einstein's summation convention is always assumed (if not
specified otherwise).  The chiral $SU(2)$ matrix $U$ has the form
$U=\exp(2i \tau_a \pi_a/F_\pi)$, where $\pi_a$ ($a=1,2,3$) are the
Cartesian isospin-components of the pion field.  $F_\pi=2f_\pi$ is the
pion-decay constant, while $e$ is the dimensionless Skyrme constant.
${\cal M}_-$ $\equiv$ $\sqrt{(m_{\pi^\pm}^2- m_{\pi^0}^2)/2}$ is
defined in terms of the masses of the charged and neutral pions,
where for the sake of convenience the mass of the neutral pion is
denoted as $m_{\pi}\equiv m_{\pi^0}$ in \re{L2}, 
\re{L_chsb} and \re{lmedend}.
Finally, $\chi_s^{00}$, $\chi_s^{02}$, $\chi_p^0$, $\Delta\chi_s$, and
$\Delta\chi_p$ represent the medium functionals which express the
influence of the surrounding environment onto the single-skyrmion
properties.  The numerical values of the input parameters in the
present work have slightly been changed in comparison to the
work~\cite{Meissner:2007ya}.  The parameters $F_\pi$ and $e$ are fixed
in such a way that the free-space (density $\rho=0$) 
PDG-values of the proton and neutron masses
($m_{\rm p}=938.27$~MeV and $m_{\rm n}=939.56$~MeV~\cite{Yao:2006px})
together with 
the empiri\-cal (isospin-averaged) mass of the delta ($m_{\Delta}
= 1232$~MeV) are
reproduced\footnote{Note that in the previous
  work~\cite{Meissner:2007ya} the input for $F_\pi$ and $e$ was
  determined by a fit to the {\em isospin-averaged} masses of both,
  the nucleon and the $\Delta$.}.  Furthermore, as input for the free
mass of the neutral pion we still use the PDG-value~\cite{Yao:2006px}:
$m_{\pi^0}=134.977$~MeV. All of these choices together induce the
following values for the input-parameters: $F_{\pi}= 108.783$~MeV, $e
= 4.854$, and  $m_{\pi^\pm}=135.015$~MeV; the latter 
is the strong part of the mass
of the charged pion.  Note that the dominant electromagnetic
contribution to $m_{\pi^\pm}-m_{\pi^0}$ is still beyond the scope of
the model.

In the Lagrangian~\re{lmedbeg}-\re{asym} the term 
${\cal L}^*_{\rm  sym}$ expresses 
the isospin-symmetric part, whereas $\Delta{\cal  L}_{\rm mes}$ 
and $\Delta{\cal L}^*_{\rm env}$ are the
isospin-brea\-king terms that arise from the explicit symmetry
breaking in the mesonic sector and the isospin asymmetry of the
surrounding environment, respectively.

\subsection{Medium functionals}
\label{med_func}

The medium functionals are given as in~\cite{Meissner:2007ya}: 
\begin{eqnarray}
\chi_{s}^{00}&=&
\left(\tilde b_{0}+\dsf{3k_{\rm F}}{8\pi^2\eta}\tilde b_{0}^2\right)\rho,
\label{mf-beg}\\
\chi_{s}^{02}&=&
\left(\tilde b_0+\dsf{3k_{\rm F}}{4\pi^2\eta}\left(\tilde b_0^2-\tilde b_1^2\right)
\right)\rho,\\
\chi_p^0&=&\dsf{2\pi c^+}{1+4\pi g^\prime c^+}
\!+\!\dsf{2\pi c^-}{1+4\pi g^\prime c^-},\\
\Delta\chi_s&=&\tilde b_1\delta\rho,\\
\Delta\chi_p&=&-\dsf{2\pi m_\pi}{\eta m_{\rm N}}\, 
c_1\left(\vec\nabla^2\delta\rho\right),
\end{eqnarray}
where
\begin{eqnarray}
c^\pm&\equiv&\Big(c_0\rho\mp c_1\delta\rho\Big)\eta^{-1},\\
\rho(\vec r)&=&\rho_{\rm n}(\vec r)+\rho_{\rm p}(\vec r),\\
\delta\rho(\vec r)&=&\rho_{\rm n}(\vec r)-\rho_{\rm p}(\vec r),
\label{mf-end}
\\
\eta&=&1+m_{\pi}/m_{\rm N}.
\end{eqnarray}
Here $k_{\rm F}=[3\pi^2\rho/2]^{1/3}$ is the total Fermi momentum,
$\rho_{\rm n}(\vec r)$ and $\rho_{\rm p}(\vec r)$ are the neutron- and
proton-distribution densities in the nucleus, 
$m_{\rm N}$ is the isospin-averaged
nucleon mass: $m_{\rm N}=(m_{\rm p}+m_{\rm n})/2$.  The parameters of
the medium functionals have the values $\tilde
b_0=-1.206m_{\pi}^{-1}$, $\tilde b_1=-1.115m_{\pi}^{-1}$,
$c_0=0.21m_\pi^{-3}$, $c_1 =0.165m_\pi^{-3}$ and
$g^\prime=0.47$~\cite{Meissner:2007ya}.  The relations between the
parameters $\tilde b_{0,1}$ and the $s$-wave pion nucleon scattering
lengths $b_{0,1}$, based on chiral perturbation theory at order ${\cal
  O}(m_\pi^3)$, can be found in
refs.~\cite{Kaiser:2001bx,Kolomeitsev:2002gc}.  As input for the
$p$-wave scattering volumes, $c_0$ and $c_1$, we use the threshold
values of the `current' SAID analysis~\cite{SAID}\footnote{For more
  references and explanations about the chosen values of $\tilde
  b_{0,1}$ and $c_{0,1}$ see ref.~\cite{Meissner:2007ya}.}.

\subsection{Proton and neutron distributions in finite nuclei}
\label{subsect:dd}

{}From now on  we will only concentrate on nuclei in the ground state and 
furthermore consider only nuclei which either are ``magic'' or which
are very near to shell closure. This allows us to use
spherically symmetric approximations to the nucleon distributions 
inside the nuclei. Consequently, 
the  distribution densities of protons and neutrons in finite nuclei 
are parameterized in the standard way~\cite{Akhiezerbook} as
\begin{eqnarray}
\rho_{\rm p,n}(r)&=&(Z,A-Z)\dsf{3}{4\pi r_{\rm p,n}^{\prime3}}\left(1
+\dsf{\pi^2a_{\rm p,n}^{\prime2}}
{r_{\rm p,n}^{\prime2}}\right)^{-1}\no\\
&&\mbox{}\times\dsf{1}{1+\exp\{(r-r_{\rm p,n}^\prime)/a_{\rm p,n}^\prime\}}\,,
\label{rho}
\end{eqnarray}
in terms of  the mass-number- and isospin-dependent parameters
\begin{eqnarray}
r_{\rm p,n}^\prime&=& r^{(0)}_{\rm p,n}A^{1/3}+r^{(1)}_{\rm p,n}
+r^{(2)}_{\rm p,n}\lambda,\qquad\no\\
a_{\rm p,n}^{\prime}&=&a^{(1)}_{\rm p,n}+a^{(2)}_{\rm p,n}\lambda\,.
\label{iso-dep-par}
\end{eqnarray}
In these expressions
$A$ is the mass number,
$Z$ is the number of protons, and 
$\lambda=(A-2Z)/A$ is the isospin-asymmet\-ry parameter.
As input for the parameters of the neutron- and proton-distributions 
we take the values
\[
\begin{array}{lcrclcr}
r^{(0)}_{\rm p}&=& 1.2490~{\rm fm},&\quad& r^{(0)}_{\rm n}&=& 1.2131~{\rm fm},\\
r^{(1)}_{\rm p}&=&-0.5401~{\rm fm},&& r^{(1)}_{\rm n}&=&-0.4415~{\rm fm},\\
r^{(2)}_{\rm p}&=&-0.9582~{\rm fm},&& r^{(2)}_{\rm n}&=& 0.8931~{\rm fm},\\ 
a^{(1)}_{\rm p}&=& 0.4899~{\rm fm},&& a^{(1)}_{\rm n}&=& 0.4686~{\rm fm},\\ 
a^{(2)}_{\rm p}&=&-0.1236~{\rm fm},&& a^{(2)}_{\rm n}&=& 0.0741~{\rm fm}.
\end{array}
\]
These values were 
obtained from calculations based on 
effective Skyrme interactions used in the description
of  the properties of finite nuclei~\cite{Hofmann:1997zu}.

As our main interest is  the 
Nolen-Schiffer anomaly, we will concentrate our investigations to 
the following pairs of mirror 
nuclei: $^{15}$N versus 
$^{15}$O, $^{17}$O versus  $^{17}$F, $^{39}$K versus  $^{39}$Ca, 
and finally $^{41}$Ca versus $^{41}$Sc. 
An important point in the calculation of the effective nucleon properties
is the proper definition of the nuclear background density. 
For example, by adding either a  neutron or a proton to the background of the 
spherical nucleus $^{16}$O one can get the nucleus  
$^{17}$O or  $^{17}$F, respectively. 
We assume that possible changes in the structure 
of the $^{16}$O core due to the additional valence nucleon are small 
and can be ignored.

\section{Parameterization of the classical solutions  I
(spherically symmetric configurations)}
\label{sect:parI}

When the skyrmion is located at the center of the nucleus, 
one can still use the spherically symmetric hedgehog ansatz
\begin{equation}
U=\exp\left[i\vec{\tau}\cdot (\vec{r}/r) F(r)\right]\,
\end{equation}
and, following the two-step method of
refs.~\cite{Meissner:2007ya,Meissner:2006id}, construct the
time-dependent Lagrangian in terms of the standard angular velocities
$\omega_i$ of the collective modes and the const\-rained angular
velocity $a^*$.  Then the time-dependent Lagrangian is given as
\begin{eqnarray}
L^*&=&\int\! {\cal L}^* {\rm d}^3 {r}
=-M_{\rm NP}^*\!-\!{\cal M}_-^2\Lambda_-\!+\!
\dsf{\vec\omega^2}{2}\Lambda^*\no\\
&+&\!\omega_3\big(a^*\Lambda^*\!+\!\Delta^*\big)\!+\!
a^*\!\left(\!\dsf{a^*}{2}\Lambda^*\!+\!\Delta^*\!\right),
\label{lag2}
\end{eqnarray}
where 
\begin{eqnarray}
M_{\rm NP}^*&=&\pi\int\limits_0^\infty\bigg\{
\dsf{F_\pi^2}{2}\left(1-\chi_{p}^0\right)\left(F_r^2 
+\frac{2\,S^2}{r^2}\right)
\nonumber\\
&+&
\dsf{2}{e^2}\left(2F_r^2+\frac{S^2}{r^2}\right)\frac{S^2}{r^2}
\\
&+&
F_\pi^2\left(m_\pi^2+{\chi_{s}^{00}}\right)
\left(1\!-\!\cos F\right)\!\bigg\}r^2{\rm d}r,\no
\end{eqnarray}
is the in-medium mass of the solution when it is not perturbed (NP) 
by any isospin-breaking effects in the mesonic sector or the 
nuclear environment. The abbreviations $F_r\equiv {\rm d} F/{\rm d} r$ and 
$S\equiv\sin F$ have been used, where $F=F(r)$ is the chiral profile function
of the hedgehog ansatz. 

\begin{eqnarray}
\Lambda_-&=&\dsf{2\pi}{3}F_\pi^2\int\limits_0^\infty S^2\,
   r^2\,{\rm d}r\
\end{eqnarray}
is a moment-of-inertia-type term resulting 
from the explicit (strong) isospin-breaking due to the
pion masses, 
whereas
\begin{eqnarray}
\Lambda^*\!&=&\!
\dsf{2\pi}{3}F_\pi^2\int\limits_0^\infty \big(1\!+\!m_{\pi}^{-2}\chi_s^{02}\big)
S^2 r^2{\rm d}r\no\\
&+&\dsf{8\pi}{3e^2}\int\limits_0^\infty 
\left(F_r^2+\frac{S^2}{r^2}\right)S^2 \,r^2\,{\rm d}r\,
\end{eqnarray}
is the moment of inertia  of the skyrmion in the nuclear medium. Finally, 
\begin{equation}
\Delta^*=\dsf{2\pi}{3}F_\pi^2\int\limits_0^\infty \Delta\alpha\,S^2\,
   r^2\,{\rm d}r
\label{Delta}
\end{equation}
with
\begin{equation}  
\Delta\alpha =\dsf{1}{2m_\pi}\left(\Delta\chi_s+\Delta\chi_p\right)
\label{Dalpha}
\end{equation}
is the isospin-breaking factor due to the isospin asymmetry of the
nuclear environment.  We note that on top of  the terms considered
in the previous refs.~\cite{Meissner:2007ya,Meissner:2006id} 
also  density gradients of the surrounding
environment are relevant which result from  
the $p$-wave pion-nucleus scattering.

In order to recover the minimization functional in the classical
approximation ($\omega_i\equiv 0$), 
one applies the constraint~\cite{Meissner:2006id}
\begin{equation}
a^{*2}=2{\cal M}_-^2{\Lambda_-}/{\Lambda^*}\,
\label{constr}
\end{equation}
such that from eq.~\re{lag2} the  Lagrangian
\begin{equation}
L^*=-M_{\rm NP}^*+a^*\Delta^*
\label{lclassic}
\end{equation}
is generated for vanishing $\omega_i$.
The variation of the Lagrangian~\re{lclassic}
gives the pertinent equation of motion for the hedgehog profile 
function $F(r)$,
\begin{eqnarray}
&&F_\pi^2(1-\chi_p^0)
\left(r^2F_{rr}+{2}{r}F_r-{S_2}\right)
-F_\pi^2\chi_{p,r}^0r^2F_r\no\\
&&+\dsf{4}{e^2}\left[{2S^2}F_{rr}+{S_2}
\left(F_r^2-\dsf{S^2}{r^2}\right)\right]
\label{classic-eq}
\\
&&-{F_\pi^2}\left(m_{\pi}^2+\chi_s^{00}\right)Sr^2
+\dsf{2a^*F_\pi^2}{3}\Delta\alpha
S_2r^2=0\,,\no
\end{eqnarray}
where the additional abbreviations $F_{rr}={\rm d}^2 F/{\rm d}r^2$,
$S_2=\sin 2 F$ and $\chi_{p,r}^0={\rm d}\chi_p^0/{\rm d}r$ 
were introduced. 
The boundary conditions
\begin{eqnarray}
\lim_{r\rightarrow 0}F(r)\!&=&\!\pi-C r\,,\no\\
\lim_{r\rightarrow \infty}F(r)\!&=&\!{D}\left(1+m_\pi r\right)
\dsf{e^{-m_\pi r}}{r^2}\,,
\label{infasymp}
\end{eqnarray}
where $C$ and $D$ are constants,
correspond to a classical soliton 
of baryon number $B=1$.

As an alternative, we will also use  the following trial function 
for approximating 
the exact solutions of the above-given differential equation~\re{classic-eq}: 
\begin{eqnarray}
F(r)&=&2\arctan\left\{\dsf{r_0^2}{r^2}(1+m_\pi r)\right\}
e^{-f(r)r},\no\\
f(r)&=&\beta_0+\beta_1e^{\,\beta_2r^2}\,.
\label{trfunc}
\end{eqnarray}
Here $r_0$, $\beta_{0}$,
$\beta_{1}$, and $\beta_{2}$ are variational parameters. Note that our trial 
function $F(r)$ differs from the widely used trial function 
$2\arctan\left\{{r_0^2}/{r^2}\right\}$, 
since it takes into account not only the asymptotical behavior at the origin
and the far-distance region, 
but approximates also the behavior in the intermediate
range.
The trial ansatz 
will be useful in our following studies of approximate solutions 
in the nonspherical  case.

As in the previous work~\cite{Meissner:2007ya} 
the final value of  $a^*$ 
is found by  an iteration procedure, starting at $a^*=0$.

Whenever the skyrmion 
can be parameterized as a spherically symmetric hedgehog configuration,
the quantization procedure and the expressions for the electromagnetic part 
of the neutron-proton mass difference remain the same as given in
in  ref.~\cite{Meissner:2007ya}, except for the fact 
that some global medium functionals
acquire a local density dependence (see eq.s~\re{mf-beg}-\re{mf-end})
and that the expression of 
$\Delta^*$  also has an additional 
contribution from $p$-wave pion-nucleus 
scattering\footnote{Compare the present expressions~\re{Delta} and~\re{Dalpha}
with the expressions (40) and (38) of ref.~\cite{Meissner:2007ya}.}.

\section{Results in the spherically symmetric approximation}
\label{sect:res_sym}

The results arising from (i)   
the direct solutions of equations of motion~\re{classic-eq} 
 and (ii) from the approximate variational 
procedure with the trial function~\re{trfunc} are presented in Table~\ref{tab1}
for several nuclei. 
%%%%%%%%%%%%%%%%%%%%%%%%%%% Begin of table 1 %%%%%%%%%%%%%%%%%%
\begin{table*}
\caption{\label{tab1} 
The values of the variational parameters and the
static properties of nucleons when the pertinent 
skyrmion  is either in free space or added to the center of a 
finite (background)  nucleus. 
Here $m_{\rm p}^*$ is the in-medium proton mass, $\Delta m_{\rm np}^*$ is 
the in-medium neutron-proton mass difference 
 and $\Delta m_{\rm np}^{*\rm (EM)}$ is its electromagnetic part; 
$\mu_{\rm p}^*$ and  $\mu_{\rm n}^*$ are the in-medium
proton and neutron
magnetic moments in units of nucleon Bohr magnetons (n.m.) in free space; 
$\langle r^2\rangle^{*1/2}_{\rm E,S}$ and $\langle r^2\rangle^{*1/2}_{\rm E,V}$
are the in-medium isoscalar (S) and isovector (V) charge radii of the nucleons.
For a given nucleus 
we present (i)~the results corresponding to the exact solution
of the differential
equation~\re{classic-eq} and (ii)~the
results of the approximate variational procedure where 
 the trial function~\re{trfunc} has been used.
}
%\footnotesize\rm
\begin{tabular*}{\textwidth}{@{}lcrrrr*{15}{@{\extracolsep{0pt plus12pt}}l}}
\hline\hline\noalign{\smallskip}
\multicolumn{2}{c}{\quad Element}&\qquad$r_0$\,\,
&\quad$10\beta_0$&\quad$\beta_1$\,\,
&\quad$\beta_2$\,\,&\,\,$m_{\rm p}^*$&$\Delta m_{\rm np}^*$&
$\Delta m_{\rm np}^{*\rm (EM)}$
&\,\,$\mu_{\rm p}^*$&\,\,$\mu_{\rm n}^*$
&$\langle r^2\rangle^{*1/2}_{\rm E,S}$&
$\langle r^2\rangle^{*1/2}_{\rm E,V}$\\
&&[fm]&[$m_\pi$]&[$m_\pi$]
&[$m_\pi^2$]
&[MeV]&[MeV]&[MeV]&[n.m.]&[n.m.]&[fm]&[fm]\\
\noalign{\smallskip}
\hline\hline
\noalign{\smallskip}
\qquad free
&(i)&-&-&-&-&938.268&1.291&-0.686&1.963&-1.236&0.481&0.739\\
\qquad space  &(ii) &0.954& 0.075&1.311&-0.009 
            &938.809&1.313&-0.687&1.966&-1.241&0.481&0.739\\
\noalign{\smallskip}
\hline
\noalign{\smallskip}
&(i)&-&-&-&-&593.285&1.668&-0.526&2.355&-1.276&0.656&0.850\\
\qquad$^{14}$N&(ii) &1.393& 0.076&0.920&0.226
            &598.505&1.655&-0.536&2.230&-1.209&0.648&0.810\\
\noalign{\smallskip}
\hline
\noalign{\smallskip}
&(i)&-&-&-&-&585.487&1.697&-0.517&2.393&-1.297&0.667&0.863\\
\qquad$^{16}$O&(ii) &1.426& 0.076&0.907&0.219
            &590.175&1.685&-0.527&2.341&-1.232&0.660&0.825\\
\noalign{\smallskip}
\hline
\noalign{\smallskip}
&(i)&-&-&-&-&558.088&1.804&-0.480&2.584&-1.422&0.722&0.942\\
\qquad$^{38}$K&(ii) &1.493& 0.076&0.841&0.153
            &559.957&1.802&-0.485&2.550&-1.377&0.718&0.910\\
\noalign{\smallskip}
\hline
\noalign{\smallskip}
&(i)&-&-&-&-&557.621&1.804&-0.478&2.569&-1.428&0.724&0.947\\
\qquad$^{40}$Ca&(ii)&1.489& 0.076&0.839&0.149
            &559.378&1.802&-0.483&2.557&-1.383&0.720&0.914\\
\noalign{\smallskip}
\hline\hline
\end{tabular*}
\end{table*}
%%%%%%%%%%%%%%% end of table 1 %%%%%%%%%%%%%%%%%%
One can see that the trial function~\re{trfunc} very well 
approximates the actual solutions of the model. The deviations from the
exactly calculated 
static properties of the  nucleon
due to the approximation~\re{trfunc} are in the range of $0.01\div5.4~\%$.
This can be seen by comparing the lines (i) and (ii) for each nucleus presented 
in Table~\ref{tab1}. In free space the deviations are even smaller:
the maximum deviation is in $\Delta m_{\rm np}$ and 
is approximately equal to $2\%$.

As expected, in accordance to our previous calculations 
for infinite nuclear matter in the 
isospin symmetric case~\cite{Meissner:2007ya}, 
the neutron-proton mass difference is slightly 
increased in comparison to the free space case.
Also note  that in this model 
the effective  nucleon mass has a strongly decreased value  
at the center of nucleus.
We will return  to this point later on.

\begin{figure}
\epsfysize=6.cm
\centerline{\epsffile{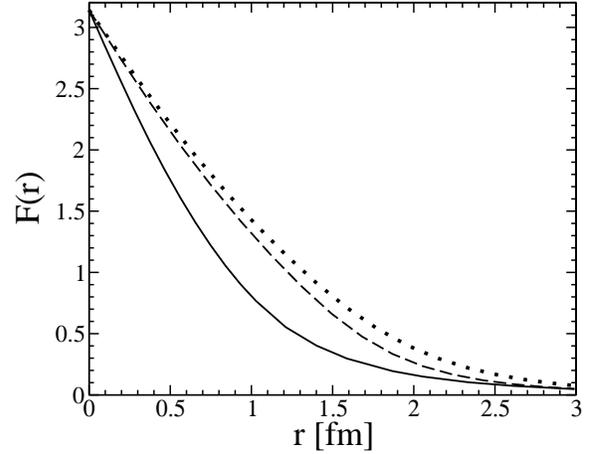}}
\vspace{0.2cm}      
\caption{The solutions of eq.~\re{classic-eq}.
The solid curve represents the profile function of a skyrmion in free space, 
the dashed curve stands  
for the profile function of a skyrmion located at the center  
of an $^{16}$O background, and the dotted curve represents the case of a  
skyrmion located 
at the center of a $^{40}$Ca core.
The approximate trial function~\re{trfunc} gives almost identical results 
(indistinguishable at the scale of the figure)
for each of these three cases.
}
\label{fig-prof}
\end{figure}

For illustrative purposes the solutions of the differential 
equation~\re{classic-eq} 
for a skyrmion in free space as well as inside  the
nuclei
 $^{16}$O and $^{40}$Ca   are shown in fig.~\ref{fig-prof}.
%%%%%%%%%%%%%%%%%%%%%%%%%% fig 1 %%%%%%%%%%%%%%%%%%%%%%%%%%%%%%%%%%%
%%%%%%%%%%%%%%%%%%%%%%% end of fig 1 %%%%%%%%%%%%%%%%%%%%%%%%%%%%%%%%%%%%%%
Note that the profile function is rather insensitive  
to small density variations: 
the solutions of the nuclei $^{14}$N and $^{16}$O 
or $^{38}$K and $^{40}$Ca, respectively,  almost coincide. 
Finally, from the results presented in Table~\ref{tab1}, it is clear 
that the trial 
function~\re{trfunc} with appropriately chosen parameters 
is almost as good  as the true solution. For this reason the trial functions
are not presented in fig.~\ref{fig-prof}, since graphically they cannot be
distinguished from their exact counterparts.

\section{Nonspherical solutions}
\label{sect:nonsph_sol}

\subsection{Time--dependent Lagrangian}
\label{subsec:tdepL}

\begin{figure}
\epsfysize=7.cm
\hskip 0.4cm \epsffile{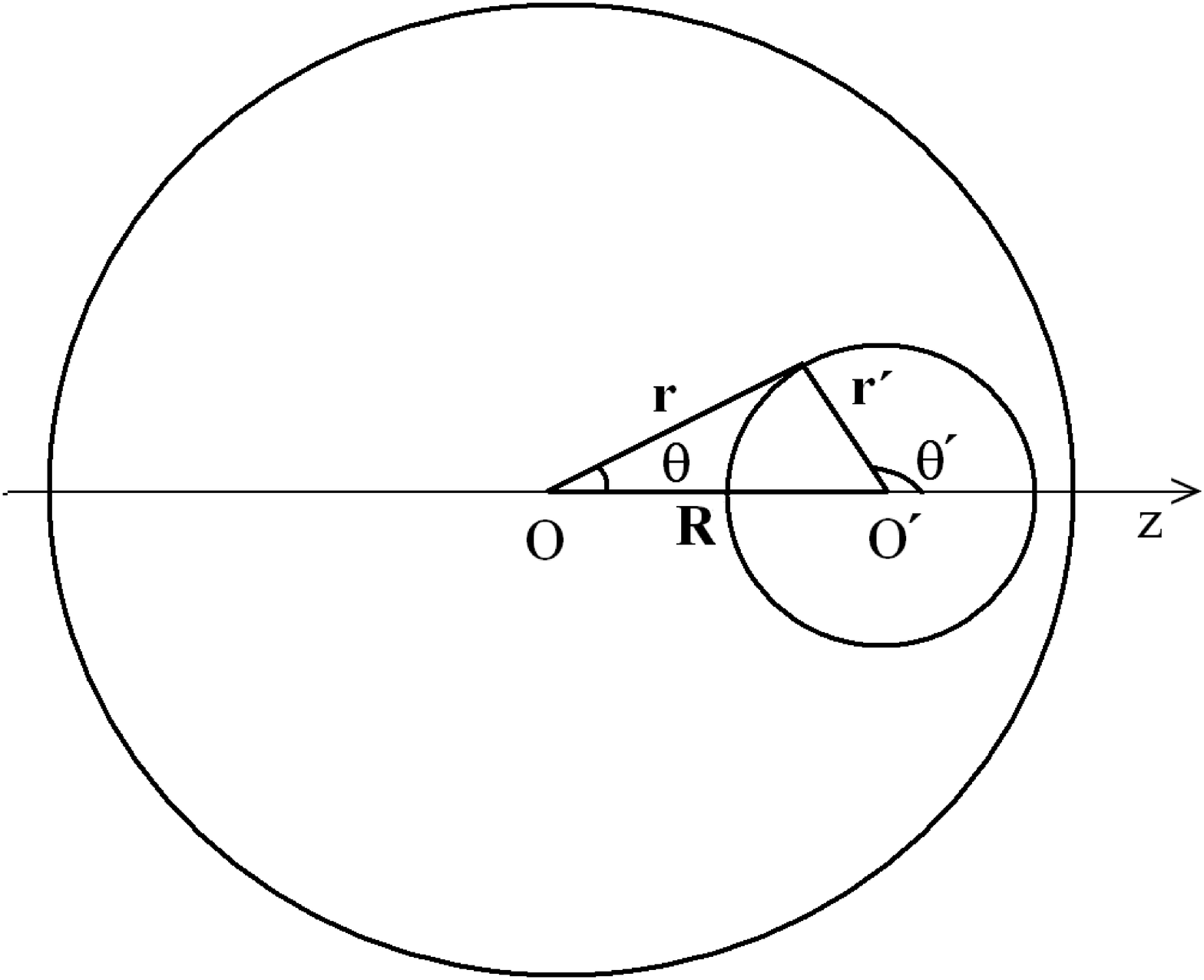}
\vspace{0.2cm}      
\caption{
A sketch of a skyrmion  located inside a finite nucleus.
Here  $\vec R$ is the distance between the center of the nucleus (O) and the 
geometrical center of  the skyrmion (O$^\prime$); 
the vectors (angles) $\vec r$ 
($\theta$) and
$\vec r'$ ($\theta'$) refer to 
the body-fixed coordinates of the nucleus and skyrmion,
respectively. Without loss of generality both coordinate systems can be 
orientated in such a way that their $z$-axes  coincide.}
\label{geometry}
\end{figure}

When the skyrmion is located at some  finite distance from the center of the
nucleus, the spherically symmetric hedgehog ansatz cannot be used
anymore, since the background profile is not spherically symmetric. Thus
there exist further  deformations in the isotopic as well as in the coordinate  
space~\cite{Yakhshiev:2001ht,Yakhshiev:2002sr}.  In this case one
should use a more generalized  ansatz  for the matrix $U$:
\begin{eqnarray}
U(\vec r\!-\!\vec R)&\!=\!
& \exp\left[i\vec\tau\!\cdot\!\vec N\!\left(\vec r\!-\!\vec R)
P(\vec r\!-\!\vec R\right)\right]\,,
\label{Proff}
\end{eqnarray}
where $\vec R$ is the distance between the geometrical centers of the
skyrmion and the nucleus (see fig.~\ref{geometry}).
%%%%%%%%%%%%%%%%%%%%%%%%%% fig 2  %%%%%%%%%%%%%%%%%%%%%%%%%%%%%%%%%%%
%%%%%%%%%%%%%%%%%%%%%%% end of fig 2  %%%%%%%%%%%%%%%%%%%%%%%%%%%%%%%%%%%%%%
Furthermore note that the argument of the profile function $P$  depends on
the direction in coordinate
space\footnote{Our notation distinguishes between 
spherically symmetric profile functions ($F$) and
  direction-dependent ones ($P$).}.  For consistency, the
isotopic vector $\vec N$ must also be generalized  as 
\begin{eqnarray}
\vec N(\vec r-\vec R)&=&\left(\begin{array}{l}
\sin\Theta(\vec r-\vec R)\cos\varphi\\
\sin\Theta(\vec r-\vec R)\sin\varphi\\
\cos\Theta(\vec r-\vec R)\end{array}
\right)\,
\label{Ng}
\end{eqnarray}
in terms of a distance- and direction-dependent {\em generalized} angle 
$\Theta=\Theta(\vec r-\vec R)$.
Furthermore, due to the azimuthal symmetry of the problem
(again see fig.~\ref{geometry}), one
can restrict the form of  both functions as  
\begin{eqnarray} 
P=P(|\vec r-\vec R|,\theta), &\quad&  
\Theta=\Theta(|\vec r-\vec R|,\theta)
\label{funcdep}
\end{eqnarray}
and put the origin of the coordinate system at the geometrical center
of the skyrmion: $\vec r^\prime = \vec r-\vec R$. For notational convenience 
we rename $\vec r^\prime$ to $\vec r$, keeping in mind
that the medium functionals acquire an $\vec R$ dependence (see
appendix~\ref{app:mass_momin})\footnote{This redefinition procedure is
  explained in more detail in ref.~\cite{Yakhshiev:2001ht}.}.

In order to apply the collective-coordinate-method to the
configuration given in \re{Proff}-\re{funcdep} we consider the following 
independent
rotations in coordinate space and in isotopic space
\begin{eqnarray}
P=P\left({\cal R}^{-1}(t)\vec r\right), &\quad&  
\vec N={\cal I}(t)\vec N\left({\cal R}^{-1}(t)\vec r\right),
\end{eqnarray}
where ${\cal R}$ and ${\cal I}$ are rotational and iso-rotational matrices,
respectively.  
Applying these time-dependent rotations to 
the two-step method of ref.~\cite{Meissner:2007ya}, one
arrives at a Lagrangian which -- after a spatial integration --  is given as
\begin{eqnarray}
L^*&=&-M_{\rm NP}^*-{\cal M}_-^2\Lambda_{\rm mes}
+\dsf{\omega_1^2\!+\!\omega_2^2}{2}\Lambda_{\omega\omega,12}^{*}\no\\
&-&(\omega_1\Omega_1\!+\!\omega_2\Omega_2)\Lambda_{\omega\Omega,12}^{*}
+\dsf{\Omega_1^2\!+\!\Omega_2^2}{2}\Lambda_{\Omega\Omega,12}^{*}\no\\
&+&\dsf{(\omega_3\!-\!\Omega_3+a^*)^2}{2}\Lambda_{\omega\Omega,33}^{*}\no\\
&+&(\omega_3\!-\!\Omega_3+a^*)\Lambda_{\rm env}^{*}\,.
\label{Lag-t}
\end{eqnarray}
Here $\omega_i$ and  $\Omega_i$  are the angular velocities in 
isotopic and coordinate space, respectively. The explicit 
expressions for the functionals, 
$M_{\rm NP}^*[P,\Theta]$ and each $\Lambda[P,\Theta]$,
can be found in appendix~\ref{app:mass_momin}.

\subsection{Parameterization of the classical solutions II}
\label{subsec:parII}

As in our previous works~\cite{Meissner:2007ya,Meissner:2006id},
by incorporating the constraint for the angular-velocity parameter $a^*$, which
in  the azimuthally symmetric case  has the form
\begin{eqnarray}
a^{*2}&=&2{\cal M}_-^2\,\dsf{\Lambda_{\rm mes}}{\Lambda^*_{\omega\Omega,33}}\,,
\end{eqnarray}
one arrives at the following Lagrange functional at the classical level:
\begin{eqnarray}
L^*&=&-M_{\rm NP}^*+a^*\Lambda_{\rm env}^{*}\,.
\label{min-func}
\end{eqnarray}
In order to extremize the Lagrange functional~\re{min-func}, 
one would have  to solve the complicated system of 
coupled partial differential equations 
that arises from the variation of the Lagrangian.
To avoid  all the unnecessary 
technical difficulties connected with a numerical
solution of that equation system, we rather
use an approximate variational procedure that is specified as follows:
first of all, as fig.~\ref{fig-prof} indicates,  
the profile function is only weakly dependent 
on the density of the medium (compare the solid and dashed curves of   
fig.~\ref{fig-prof}) and almost insensitive to small density changes 
 (compare the dashed and dotted curves of fig.~\ref{fig-prof}).
Consequently, we can apply the approximations 
\begin{eqnarray}
P(r,\theta)&=&2\arctan\left\{\dsf{r_0^2}{r^2}(1+m_\pi r)
\big(1+u(\theta)\big)\right\}
e^{-f(r)r},\no\\
\Theta(r,\theta)&=&\theta+\zeta(r,\theta)\,,
\label{app1}
\end{eqnarray}
where the special case $P(r,0)=F(r)$ is the 
spherically symmetric profile function 
(in the form as presented in eq.~\re{trfunc}) and where
the functions $u$ and $\zeta$ satisfy the inequalities $|u|<1$ and $|\zeta|<1$ 
in the regions
 $r\in[0,\infty)$ and $\theta\in[0,\pi]$. 
Following the ideas of ref.~\cite{Yakhshiev:2001ht} we use 
for the function $u$ the following ansatz
\begin{eqnarray}
u(\theta)&=& \sum_{n=1}^\infty\gamma_n\cos^n\theta\,,
\label{ffunc}
\end{eqnarray}
where the $\{\gamma_n\}$ 
are variational parameters and the cosine functions are chosen 
to maintain periodicity in $\theta$.
Similarly, $\zeta$ can be chosen as
\begin{eqnarray}
\zeta(r,\theta)&=& re^{-\delta_0^2 r^2}\sum_{n=1}^\infty\delta_n\sin 2n\theta\,,
\label{zetafunc}
\end{eqnarray}
where the $\{ \delta_n\}$ are also variational parameters. Note that 
the arguments 
of the sine-functions are chosen to be a multiple of $2\theta$ in order 
to avoid singularities 
of the form $\sin\Theta/\sin\theta$ in the Lagrange functional~\re{min-func}
(see appendix~\ref{app:mass_momin} for the explicit form of 
the Lagrange functional).
Furthermore, the $r$ dependence of  $\zeta$ is chosen in such a way that
the equalities 
$\Theta(0,\theta)=\theta$ and $\Theta(\infty,\theta)=\theta$ are reproduced.

Finally, in terms of the trial functions~\re{app1}, 
\re{ffunc} and \re{zetafunc}, the Lagrange
functional~\re{min-func} will be extremized.

\subsection{Quantization of the nonspherical solutions}
\label{subsect:nonsph-sol}

Defining canonical conjugate variables in the body-fixed reference system 
\begin{eqnarray}
T_i=\dsf{\partial{L^*}}{\partial\omega_i}\mbox{\quad and\quad}
J_i=\dsf{\partial{L^*}}{\partial\Omega_i}\,,
\end{eqnarray}
one obtains from the time-dependent Lagrangian~\re{Lag-t} the Hamiltonian:
\begin{eqnarray}
\hat H&=&M_{\rm NP}^*+{\cal M}_-^2\Lambda_{\rm mes}
+\dsf{\Lambda_{\rm env}^{*2}}{2\Lambda_{\omega\Omega,33}^*}\no\\
&+&\dsf{(\hat T_1^2+\hat T_2^2)\Lambda_{\Omega\Omega,12}^*+
(\hat J_1^2+\hat J_2^2)\Lambda_{\omega\omega,12}^*}
{2(\Lambda_{\omega\omega,12}^*\Lambda_{\Omega\Omega,12}^*
-\Lambda_{\omega\Omega,12}^{*2})}\no\\
&+&\dsf{(\hat T_1\hat J_1+\hat T_2\hat J_2)\Lambda_{\omega\Omega,12}^*}
{\Lambda_{\omega\omega,12}^*\Lambda_{\Omega\Omega,12}^*
-\Lambda_{\omega\Omega,12}^{*2}}\no\\
&+&\dsf{\hat T_3^2}{2\Lambda_{\omega\Omega,33}^*}
-\left(a^*+\dsf{\Lambda_{\rm env}^*}{\Lambda_{\omega\Omega,33}^*}\right)\hat T_3\,.
\label{Ham-final}
\end{eqnarray}
By sandwiching the Hamiltonian between the appropriate bar\-yon states 
$|T,T_3;J,J_3=-T_3\rangle$ one 
determines the energy of a nucleon inside a nucleus
as 
\begin{eqnarray}
E&=&M_{\rm NP}^*+{\cal M}_-^2\Lambda_{\rm mes}
+\dsf{\Lambda_{\rm env}^{*2}}{2\Lambda_{\omega\Omega,33}^*}\no\\
&+&\dsf{\Lambda_{\Omega\Omega,12}^*+\Lambda_{\omega\omega,12}^*
-2\Lambda_{\omega\Omega,12}^*}
{2(\Lambda_{\omega\omega,12}^*\Lambda_{\Omega\Omega,12}^*
-\Lambda_{\omega\Omega,12}^{*2})}\big(T(T+1)-T_3^2\big)\no\\
&+&\dsf{T_3^2}{2\Lambda_{\omega\Omega,33}^*}
-\left(a^*+\dsf{\Lambda_{\rm env}^*}{\Lambda_{\omega\Omega,33}^*}\right) T_3\,.
\label{E-final}
\end{eqnarray}
Consequently, the strong part of the 
neutron-proton mass difference in the interior of  a nucleus takes the form
\begin{eqnarray}
\Delta m_{\rm np}^{*(\rm strong)}
&=&a^*+\dsf{\Lambda_{\rm env}^*}{\Lambda_{\omega\Omega,33}^*}\,.
\end{eqnarray}

\section{Electromagnetic part of the neutron-proton mass 
difference in nonspherically deformed states}
\label{sect:nonsph-EMff}

The electric (E) and magnetic (M) form factors of the nucleon are defined
through the expressions
\begin{eqnarray}
G_{\rm E}^* (\vec q^2)&=&\dsf12
\int {\rm d}^3r \, e^{i\vec q\cdot\vec r}j^0(\vec r)~,\nonumber\\
G_{\rm M}^* (\vec q^2)&=&\dsf{m_{\rm N}}2
\ds\int {\rm d}^3r \, e^{i\vec q\cdot\vec r}[\vec r\times \vec j(\vec r)]~,
\label{ffdef}
\end{eqnarray}
where $\vec q^2$ is the squared  momentum transfer. Furthermore, 
$j^0$ and $\vec j$ correspond
to the time and space components of the properly normalized sum
of the baryonic current $B_\mu^*$
and the third component of the isovector current $\vec V_\mu^*$ of the Skyrme
model. 

For the problem at hand, it is advantageous to expand 
the plane wave factor in \re{ffdef}\footnote{We can rotate the total system,
background nucleus and skyrmion, always such that the $z$-axis of the
reference frame 
coincides with the 
$z$-direction of the coordinate system of the skyrmion in the 
body-fixed frame.} as
\begin{eqnarray}
e^{i\vec q\cdot\vec r}&=& e^{i qr\cos\theta}=
\sum_{l=0}^\infty i^l (2l+1)P_l(\cos\theta)J_l(qr)\,,
\end{eqnarray}
in terms of Legendre polynomials $P_l$ and spherical Bessel functions $J_l$.
In this way we get the final expressions for the electromagnetic form factors
\begin{eqnarray}
G_{a}^{*b} (q^2)&=&\sum_{l=0}^\infty 
i^l(2l+1)G_{a,l}^{*b}(q^2)\,.
\end{eqnarray}
Here 
\begin{eqnarray}
G_{a,l}^{*b}(q^2)&=&
\int {\rm d}^3r\,\rho_{a}^{b}(r,\theta)P_l(\cos\theta)J_l(qr)
\end{eqnarray}
are partial form factors, 
where the label $a$ stands either for isoscalar (S) or isovector 
(V) form factors, and the label
$b$ denotes either  electric (E) or magnetic (M) form factors. 
Explicit expressions of the corresponding
charge ($\rho_{\rm E}$) and 
magnetic  ($\rho_{\rm M}$) densities are given in appendix~\ref{app:Charges}.
As actual calculations show,
the absolute values of the partial form factors 
decrease quickly with increasing 
partial wave number $l$ within the present approach.

Finally, applying the formula~\cite{Gasser:1982ap}
\begin{eqnarray}
\Delta m_{\rm np}^{*\rm (EM)}&=& 
-\dsf{4\alpha}{\pi}\int\limits_0^\infty {\rm d}q
\Big\{G_{\rm E}^{\rm S*}(\vec q\,^2)G_{\rm E}^{\rm V*}(\vec q\,^2)\no\\
&&-\dsf{\vec q\,^2}{2m_{\rm N}^2}
G_{\rm M}^{\rm S*}(\vec q\,^2)G_{\rm M}^{\rm V*}(\vec q\,^2)\Big\}\,
\label{dm-EM}
\end{eqnarray}
where $\alpha\approx 1/137$ is the electromagnetic fine
structure constant, one can calculate the medium-depen\-dent 
electromagnetic part of the 
neutron-proton mass difference.

\section{Results in the nonspherical approximation}
\label{sect:res_nonsym}

First of all let us remark that numerically the  transitions
$$
\ds\lim_{R\rightarrow 0}\Theta(r,\theta)\rightarrow 
\theta\quad\mbox{ and }\quad
\ds\lim_{R\rightarrow 0}P(r,\theta)\rightarrow F(r)
$$ 
are smooth when the nucleon `moves' to the center of the nucleus,  
and that in this case the exact solution is
reproduced with high accuracy, see Table~\ref{tab1}.
One might therefore expect that the 
trial functions~\re{app1}-\re{zetafunc} are not too far from the
the true solutions even when 
the distance between the center of the skyrmion and the center of the 
nucleus is 
not equal to zero, $R\ne 0$.

Obviously, the variational parameters $r_0$, \{$\beta_n\}$, \{$\gamma_n\}$ 
and $\{\delta_n\}$ 
depend on the distance $R$ between the centers of the skyrmion and the 
nucleus.
Note that the strength of the parameter $\gamma_n$ decreases 
with increasing index $n$.
For example, $\gamma_3$ is one order magnitude smaller than $\gamma_1$
at those distances where the deviations of the skyrmion from the 
spherical form  is maximal. 
The strength of $\delta_n$ decreases even more rapidly
with increasing $n$: $\delta_2$
is one order of magnitude smaller than $\delta_1$ 
at the above-mentioned distances
and $\delta_3$ is almost zero for all values of $R$. 
For performing high accuracy
calculations it is  therefore sufficient to retain the
$\gamma_n$ parameters  up-to-and-including $\gamma_4$ and the $\delta_n$ 
parameters up-to-and-including $\delta_2$.

\begin{figure}
\vskip 0.2cm
\hskip 0.2cm
\epsfysize=6.cm
\epsffile{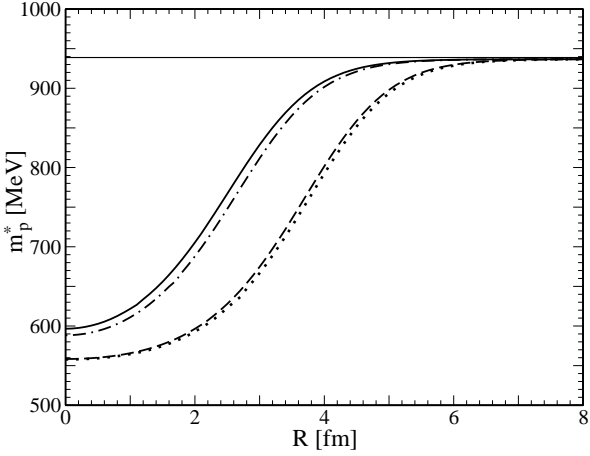}
\caption{The dependence of the effective mass of the proton 
$m_{\rm p}^*$ on the distance $R$
between the centers of the skyrmion and the nucleus. 
The solid curve represents the result
for an additional skyrmion embedded into a $^{14}$N core, 
the dot-dashed curve stands for the 
case of  $^{16}$O, 
the dashed curve refers to the case of $^{38}$K and
dotted curve refers to the case of  $^{40}$Ca.
The horizontal line indicates the pertinent
free space value of the proton mass, see line (ii) in Table~\ref{tab1}, 
{\em i.e.} $m_{\rm p}=938.81$~MeV.
}
\label{mpeff}
\end{figure}

The behavior of the static properties of the nucleons in nuclei and 
the corresponding variational parameters are discussed in a more detail in 
ref.~\cite{Yakhshiev:2001ht}. Qualitatively, 
we have similar results in the 
present approach. For example, 
the in-medium mass of the proton starts with  
the effective  value presented in Table~\ref{tab1} when it is near 
the center of the nucleus. Then it increases 
monotonically  with increasing distance 
between its center and the center of the nucleus until
it approaches the free space value at
the border of the nucleus, see fig.~\ref{mpeff}.
%%%%%%%%%%%%%%%%%%%%%%%%%% fig 3 %%%%%%%%%%%%%%%%%%%%%%%%%%%%%%%%%%%
%%%%%%%%%%%%%%%%%%%%%%% end of fig 3 %%%%%%%%%%%%%%%%%%%%%%%%%%%%%%%%%%%%%%

\begin{figure}
\vskip 0.2cm
\hskip 0.2cm
\epsfysize=6.cm
\epsffile{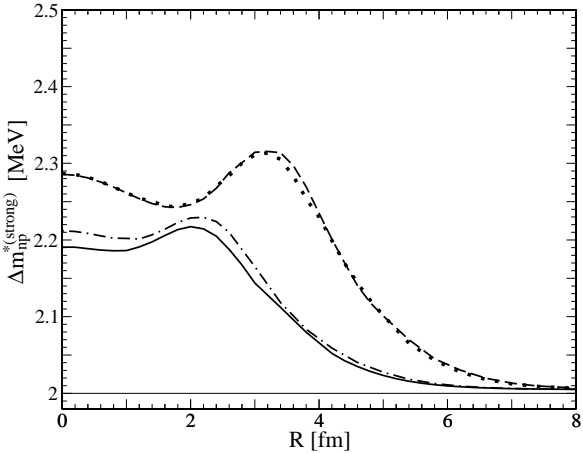}
\caption{The dependence of $\Delta m_{\rm np}^{\rm *(strong)}$ 
on the distance $R$
between the centers of the skyrmion and the nucleus. 
The notations are same as in fig.~\ref{mpeff}.
The horizontal line indicates the free space value of 
$\Delta m_{\rm  np}^{\rm (strong)}=2$~MeV (see the corresponding line (ii) in
Table~\ref{tab1}).  }
\label{strong}
\end{figure}

We will come back to the discussion of 
the effective nucleon mass in sect.~\ref{sect:NSA}.
Here, however,
we concentrate on 
the effective in-medium neut\-ron-proton mass {\em difference}.
The behavior of $\Delta m_{\rm np}^{\rm *(strong)}$ inside several 
nuclei is presented in fig.~\ref{strong}.
%%%%%%%%%%%%%%%%%%%%%%%%%% fig 4 %%%%%%%%%%%%%%%%%%%%%%%%%%%%%%%%%%%
%%%%%%%%%%%%%%%%%%%%%%% end of fig 4 %%%%%%%%%%%%%%%%%%%%%%%%%%%%%%%%%%%%%%
The strong part of the effective mass difference 
has a non-monotonic behavior. This is a consequence
of  the density  now being a local quantity 
and of the appearance of 
additional isospin-breaking contributions  due to the 
density gradients 
arising from the $p$-wave 
pion-nucleus scattering (see eq.~\re{Dalpha}).
One can explicitly see in  fig.~\ref{strong} that
at the surface of the nucleus, where the density gradients are large
and the local isospin asymmetry is high, the value of  
$\Delta m_{\rm np}^{\rm *(strong)}$ is at an extremum.

\begin{figure}
\vskip 0.2cm
\hskip 0.2cm
\epsfysize=6.cm
\epsffile{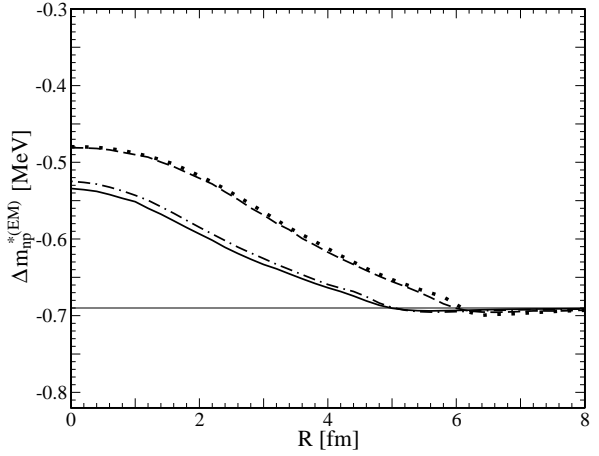}
\caption{The dependence of $\Delta m_{\rm np}^{\rm *(EM)}$ on the distance $R$
between the centers of the skyrmion and the nucleus. 
The notations are same as in fig.~\ref{mpeff}.
The horizontal line indicates the
free space value of $\Delta m_{\rm np}^{\rm (EM)}=-0.69$~MeV
(see the corresponding line (ii) in Table~\ref{tab1}).}
\label{fig-EM}
\end{figure}

In fig.~\ref{fig-EM} the electromagnetic part of the neutron-proton 
mass difference is shown for  several nuclei.
%%%%%%%%%%%%%%%%%%%%%%%%%% fig 5 %%%%%%%%%%%%%%%%%%%%%%%%%%%%%%%%%%%
%%%%%%%%%%%%%%%%%%%%%%% end of fig 5 %%%%%%%%%%%%%%%%%%%%%%%%%%%%%%%%%%%%%%
The variations in the electromagnetic part of the neutron-proton 
mass differences are small 
compared with their strong counterparts. Qualitatively, their behavior is 
the same for all nuclei as in the case of strong part. 
Quantitatively, both parts, $\Delta
m_{\rm np}^{\rm *(strong)}$ and $\Delta
m_{\rm np}^{\rm *(EM)}$, are mainly governed by the (local)
behavior of the total nuclear density and therefore differ 
if  the total densities of the investigated nuclei are different.

\begin{figure}
\vskip 0.2cm
\hskip 0.2cm
\epsfysize=6cm
\epsffile{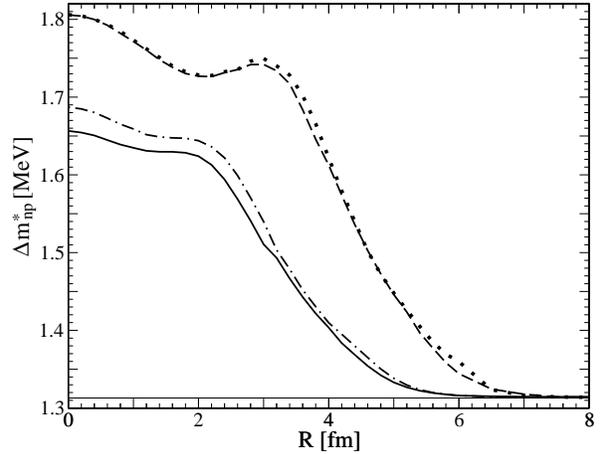}
\caption{The total 
neutron-proton mass difference  $\Delta m_{\rm np}^{*}$ in
nuclei. 
The notations are same as in fig.~\ref{mpeff}.
The horizontal line indicates the free space 
value of $\Delta m_{\rm np}=1.31$~MeV
(see the corresponding line (ii) in Table~\ref{tab1}).}
\label{total}
\end{figure}

Finally, in fig.~\ref{total} the dependence of the
total neutron-proton mass difference $\Delta m_{\rm np}^{*}$ on the distance $R$
is shown for several nuclei.
%%%%%%%%%%%%%%%%%%%%%%%%% fig 6 %%%%%%%%%%%%%%%%%%%%%%%%%%%%%%%%%%%
%%%%%%%%%%%%%%%%%%%%%%% end of fig 6 %%%%%%%%%%%%%%%%%%%%%%%%%%%%%%%%%%%%%%

\section{Calculation of the Nolen-Schiffer anomaly}
\label{sect:NSA}

A long standing problem in nuclear physics is the 
Nolen-Schiffer anomaly (NSA)  observed 
in mirror nuclei~\cite{Nolen:1969ms,Shlomo}. Let us consider 
this point in more detail
and show how the NSA can be treated in the present 
framework.

The mass difference between mirror nuclei which differ by one unit in
their charges, $Z_1-Z_2=1$, 
is usually split into two contributions
\begin{eqnarray}
\Delta M\equiv
{}^A_{Z+1}{\bf M}_N-{}^A_{Z}{\bf M}_{N+1}=\Delta E_{\rm EM}-\Delta m_{\rm np}^*,
\label{mass-dif}
\end{eqnarray}
where $\Delta E_{\rm EM}$ is the computed Coulomb energy difference,
including various corrections (by, {\em e.g.}, the 
exchange term, the center-of-mass motion, 
finite size effects of the proton and neutron charges, 
magnetic interactions,  
vacuum effects, 
the dynamical effect 
of the neutron-proton mass difference, and short-range two-body
correlations). Furthermore, the quantity  $\Delta m_{\rm np}^*$ is the
effective neutron-proton mass difference.  The measurement of the mass
difference $\Delta M$ and the theoretical calculation of 
$\Delta E_{\rm  EM}$ can be done with great accuracy (within 1\%
error)~\cite{Shlomo}.  Furthermore if one assumes that
$\Delta m_{\rm  np}^*$ is constant and equal to its free space value, then
eq.~\re{mass-dif} will not be satisfied.  This phenomenon is called
the Nolen-Schiffer anomaly, persistent throughout the periodic table.
Quantitatively, the NSA ranges between few hundred keV for the lightest nuclei
up to one MeV or more for the heaviest nuclei. A possible resolution of the
anomaly is the assumption that the effective neutron-proton mass difference
$\Delta m_{\rm np}^*$ would decrease with increasing mass number $A$.
Consequently, the NSA would be 
$\Delta_{\rm  NSA}=\Delta m_{\rm np}-\Delta m_{\rm np}^*$.

Within the present approach $\Delta m_{\rm np}^*$ has a local $R$ dependence 
according to the location of the nucleons inside the nuclei as represented 
in fig.~\ref{total}. 
In order to compare our results 
with the experimental data one has to average the value of 
 $\Delta m_{\rm np}^*$ with respect to the separation $R$.
In this respect 
we note that the nucleons inducing the Nolen-Schiffer anomaly are valence
nucleons which, if the mirror nuclei differ by one particle or hole from
the (magic) closed-shell case, must be 
located in the peripheral region of each of the mirror nuclei. 
For example, in a nuclear shell model with a 
Wood-Saxon potential
the wave functions of the valence nucleon of a nucleus of mass number 
$A=A_{\rm magic}\pm 1$, respectively,
can be fitted very well  by the Gaussian form
\begin{eqnarray}
\psi_A^{(\rm n,p)}(R)
&=&\left(\dsf{\big(b_A^{(\rm n,p)}\big)^3\big(b_A^{(\rm n,p)}R\big)^{2n_A}}
{2\pi\Gamma(3/2+n_A)}\right)^{1/2}\no\\
&\times&\exp\left\{-\dsf{\big(b_A^{(\rm n,p)}R\big)^2}{2}\right\}\,,
\label{WF}
\end{eqnarray}
where $\Gamma$ is the Gamma function and the parameters $b_A$ and $n_A$ have 
the following 
values\footnote{For the values of the Wood-Saxon potential parameters
see ref.~\cite{Chepurnov67}.}:
\begin{eqnarray}
b_{15}^{\rm (p)}=0.871m_\pi, \qquad b_{15}^{\rm (n)}=0.886m_\pi, 
\qquad n_{15}=1,\no\\
b_{17}^{\rm (p)}=0.843m_\pi, \qquad b_{17}^{\rm (n)}=0.860m_\pi, 
\qquad n_{17}=2,\no\\
b_{39}^{\rm (p)}=0.785m_\pi, \qquad b_{39}^{\rm (n)}=0.804m_\pi, 
\qquad n_{39}=2,\no\\
b_{41}^{\rm (p)}=0.774m_\pi, \qquad b_{41}^{\rm (n)}=0.794m_\pi, 
\qquad n_{41}=3.\no
\end{eqnarray}
The averaged neutron-proton-mass difference is given by\footnote{Note 
that the wave functions~\re{WF} are properly normalized.}
\begin{eqnarray}
\Delta\overline{m}^*_{\rm np}&=&\overline{m}^*_{\rm n}-\overline{m}^*_{\rm p}\,,
\label{loc_mass_dif}\\
\overline{m}_{\rm n,p}^*&\equiv&\int 
m_{\rm n,p}^*(R)\big(\psi^{\rm(n,p)}(R)\big)^2{\rm d}^3R\,.
\label{eff_mass}
\end{eqnarray} 
Defining the difference of neutron-proton probability densities as 
$$
\big(\psi^{\rm(n)}(R)\big)^2-\big(\psi^{\rm(p)}(R)\big)^2\equiv
\Delta\psi_{\rm np}^{(2)}(R)
$$
one can rewrite  eq.~\re{loc_mass_dif} as
\begin{eqnarray}
\Delta\overline{m}_{\rm np}^*&\approx&
\int\left(\Delta\psi^{(2)}_{\rm np}m_{\rm p}^* 
+\big(\psi^{(p)}\big)^2\Delta m_{\rm np}^*\right){\rm d}^3R\no\\
&\equiv&\Delta\overline{m}_{\rm np}^{*(1)}
+\Delta\overline{m}_{\rm np}^{*(2)}\,,
\label{contribs}
\end{eqnarray} 
 ignoring  the
subleading contribution of  the cross term 
$\int \Delta\psi_{\rm np}^{(2)}\Delta m_{\rm np}^*\,{\rm d}^3R$.

Then the Nolen-Schiffer anomaly for a given pair of mirror nuclei with
the mass number $A$ simply takes the form
\begin{equation}
\overline{\Delta}_{\rm NSA}=
\Delta m_{\rm np}
-\left(\Delta\overline{m}_{\rm np}^{*(1)}
+\Delta\overline{m}_{\rm np}^{*(2)}\right)\,.
\label{tildeNSA}
\end{equation}

The dependence of the Nolen-Schiffer anomaly on 
the mass number $A$ is presented in 
Table~\ref{tab2} (see  the third column 
`$\alpha_{\rm ren}$=$0$, present approach').
%%%%%%%%%%%%%%%%%%%%%%%%%%% Begin of table 2 %%%%%%%%%%%%%%%%%%
\begin{table*}[hbt]
\caption{\label{tab2} 
The averaged mass of the valence proton in a 
given nucleus $\overline{m}_{\rm p}^*$,
the contributions to the effective neutron-proton mass difference 
(see eq.~\re{contribs}) and 
the comparison of the 
Nolen-Schiffer discrepancy  $\overline{\Delta}_{\rm NSA}$ (see 
eq.s~\re{tildeNSA},\,\re{toy_par})
calculated in the present approach with other 
``empirical'' results. All quantities are in units of MeV.
}
%\footnotesize\rm
\begin{tabular*}
{\textwidth}{@{}ll|cc|ccc|ccc|c|c*{15}{@{\extracolsep{0pt plus12pt}}l}}
\hline\hline\noalign{\smallskip}
\quad\qquad&&&&
\multicolumn{6}{|c|}{Present approach}&&\\
\noalign{\smallskip}
\cline{5-10}
\noalign{\smallskip}
&Nuclei&\multicolumn{2}{|c|}{$\overline{m}_{\rm p}^*$}
&\multicolumn{3}{c|}{$\alpha_{\rm ren}=0$}&
\multicolumn{3}{c|}{$\alpha_{\rm ren}=0.95$}&$\overline{\Delta}_{\rm NSA}$
&$\overline{\Delta}_{\rm NSA}$\\
\noalign{\smallskip}
\cline{3-10}
\noalign{\smallskip}
&&$\alpha_{\rm ren}=0$&$\alpha_{\rm ren}=0.95$
&$\Delta\overline{m}_{\rm np}^{*(1)}$&$\Delta\overline{m}_{\rm np}^{*(2)}$
&$\overline{\Delta}_{\rm NSA}$&
$\Delta\overline{m}_{\rm np}^{*(1)}$&$\Delta\overline{m}_{\rm np}^{*(2)}$
&$\overline{\Delta}_{\rm NSA}$
&ref.~\cite{Nolen:1969ms}&ref.~\cite{Shlomo}\\
\noalign{\smallskip}
\hline
\noalign{\smallskip}
&$^{15}$O-$^{15}$N&767.45&928.30&-4.27 & 1.56 &4.02&-0.21&1.33&0.20& -
&  $0.16\pm0.04$\\
&$^{17}$F-$^{17}$O&812.35&930.54&-5.53 & 1.52 &5.33&-0.28&1.32&0.27&0.31 
&$0.31\pm0.04$\\
&$^{39}$Ca-$^{39}$K&724.78&926.16&-8.11& 1.67 &7.75&-0.41&1.33&0.37&-
&  $0.22\pm0.08$\\
&$^{41}$Sc-$^{41}$Ca&771.71&928.51&-9.74&1.62 &9.44&-0.49&1.33&0.47&0.62 
& $0.59\pm0.08$\\
\noalign{\smallskip}
\hline\hline
\end{tabular*}
\end{table*}
%%%%%%%%%%%%%%% end of table 2 %%%%%%%%%%%%%%%%%%
One can see that  our predictions
of the  behavior of the NSA 
qualitatively go into the right
direction. But quantitatively, the 
results are too big by one order of magnitude.  
This is mainly due to the strong shift of 
$\Delta \overline{m}_{\rm p}^{*(1)}$ 
(see the third column of   Table~\ref{tab2}) 
that results for three reasons: (i)
the rather large renormalization of the effective nucleon mass, (ii) 
the pronounced $R$ dependence of $m_{\rm p}^*$ inside the nucleus 
(see fig.~\ref{mpeff}), and
(iii)  the relative swelling of the proton distributions due to the 
Coulomb factor, {\em i.e.} $\Delta\psi^{(2)}_{\rm np}\ne 0$.  
For example, the averaged in-medium
mass of the valence proton in $^{17}$O is reduced to
$\overline{m}_{\rm  p}^*=812.35$~MeV. This drop of about  125~MeV is
very large in comparison with the
empirical value of the binding energy per
nucleon in nuclear matter.    
For heavier nuclei, where the density in the interior approximates 
the normal nuclear matter density (see fig.~\ref{mpeff}), 
the drop of the averaged effective mass is even
larger, {\em e.g.} $m_{\rm p}-\overline{m}_{\rm p}^*\sim 150$ -- 200~MeV 
around $^{40}$Ca (see the second column of Table~\ref{tab2}) down to
$\sim 300$~MeV around $^{208}$Pb.

If only the contribution 
$\Delta \overline{m}_{\rm  np}^{*(2)}$ (due to the explicit $R$ 
dependence of the
neutron-proton mass difference) were considered, 
then the NSA in the present approach
would even have a  negative sign: 
$\Delta m_{\rm np}-\Delta \overline{m}_{\rm  np}^{*(2)}<0$.

In our model the effective mass of the nucleon  strongly depends on 
the  phenomenological input parameters of the pion-nucleus sector, 
mainly via the
$p$-wave scattering volume $c_0$ and the Landau parameter $g^\prime$ (see
subsection~\ref{med_func}).  In ref.~\cite{Rakhimov:1996vq},
according to the systematics of the effective axial coupling $g_A^*$
in finite nuclei of the range $5\le A\le 39$~\cite{Buck:1975ae}, 
the values $c_0=0.15m_\pi^{-3}$ and
$g^\prime=0.6$ were predicted\footnote{Compare 
with the values $c_0=0.21m_\pi^{-3}$
  and $g^\prime=0.47$ used in the present work.}. 
These values imply
a smaller drop of the effective mass of the nucleon, {\em e.g.}
$m_{\rm N}^*/m_{\rm N}\sim 0.76$ in normal nuclear
matter~\cite{Rakhimov:1996vq}.  Although, the use of
smaller values of
$c_0$ and bigger values of $g^\prime$ improves the 
NSA results ({\em e.g.}, $\overline{\Delta}_{\rm NSA}=2.86$~MeV for
the pair $^{15}$O-$^{15}$N for $c_0=0.15m_\pi^{-3}$ and $g^\prime=0.75$),
this is still not enough to satisfy the experimental data.  

Instead of fine-tuning these parameters to unphysical values, one
might rather invert the problem and 
try to estimate
the effective nucleon mass inside finite nuclei according to the
results in the isospin-breaking sector.  To perform this task we fine-tune
an artificially added  
renormalization parameter $\alpha_{\rm ren}$ in the expression 
\begin{eqnarray}
m_{\rm n,p}^*(R,\alpha_{\rm ren})&=&m_{\rm n,p}^*(R)
+\big(m_{\rm n,p}-m_{\rm n,p}^*(R)\big)\alpha_{\rm ren}\qquad
\label{toy_par}
\end{eqnarray}
of the effective nucleon mass in such a way that
the NSA is satisfied. The results are presented in
Table~\ref{tab2} (see the column `$\alpha_{\rm ren}=0.95$').  
It can be seen that
a successful  description of the correct order of the NSA implies 
a rather small drop of the
mass of the valence nucleons: 
$m_{\rm n,p}-\overline{m}_{\rm n,p}^*(\alpha_{\rm ren}=0.95)\sim 10$~MeV 
which is close to the
empirical binding energy per nucleon in nuclear matter.  
In this case the
contribution to the NSA from $\Delta \overline{m}_{\rm np}^{*(2)}$ 
can be neglected:
$\Delta m_{\rm np}- \Delta\overline{m}_{\rm np}^{*(2)}(\alpha_{\rm ren}=0.95)
\sim -0.02$~MeV.

\section{Conclusions and Outlook}
\label{sect:sum}

In summary, we have  studied the effective neut\-ron-proton mass
difference $\Delta m_{\rm np}^*$ in finite nuclei in the framework of
a medium-modified Skyrme model.  The value of $\Delta m_{\rm np}^*$
for a given nucleus approaches an extremum near the nuclear surface
where the local density gradients are large and the isospin asymmetry
is high.

We have discussed the relevance of our results for the Nolen-Schiffer
anomaly. Qualitatively, our approach predicts the correct behavior of
the Nolen-Schiffer anomaly. But quantitatively it is not satisfactory:
the results are one order of magnitude too large. Clearly, the part of our
calculation relevant to the Nolen-Schiffer anomaly depends on the
proton and neutron distributions of the mirror nuclei and is very
sensitive to the behavior of the wave functions of the valence
nucleons in the peripheral region of the nucleus. 
We have pointed out the possibility that
the Nolen-Schiffer anomaly may rather follow from
the behavior of the effective nucleon mass 
in finite nuclei than from the effective neutron-proton mass difference:
our calculations imply  that the Nolen-Schiffer anomaly 
could not and, maybe, should not be 
saturated by $\Delta \overline{m}_{\rm np}^{*(2)}$ (the contribution due to the
explicit density dependence of the neutron-proton mass difference).  
Rather more important is   
$\Delta \overline{m}_{\rm np}^{*(1)}$, the contribution 
due to the difference in the wave functions of valence
proton and neutron  weighted by  
the local  (density and density-gradient induced) 
variation of the effective mass of the nucleon. 
In fact, when we restrict the in-medium reduction
of the (averaged effective) proton 
mass to about 1\,\% of the free proton mass, 
we obtain a rather precise description of the NSA.

Of course, the calculations of $\Delta E_{\rm EM}$  in ref.~\cite{Shlomo} 
(see eq.~\re{mass-dif}) takes into account 
the effect due to the different wave functions
of valence nucleons\footnote{This  phenomenon is known as Thomas-Ehrman
  effect~\cite{Thomas51,Ehrman51}.}. But in  ref.~\cite{Shlomo}  
a constant value of the nucleon mass was used in the calculations, 
namely the free mass. So 
it might be possible to resolve
the anomaly  of the mirror nuclei by considering  a
dynamical (local) mass of the nucleon that is only slightly reduced in 
comparison to the free mass.
In this respect, it is interesting to note that in the 
local-density-functional approach to many-body nuclear systems an 
additional term has been
introduced into the Coulomb part of the energy density of the nuclear system 
that removes the anomaly~\cite{Bulgac96,Fayans98,Bulgac99,Fayans01}.
Moreover, this term is chosen such that
it is proportional to the  {\em isoscalar} rather than the 
{\em isovector} density\footnote{Note, however, that the isoscalar 
contribution  is surface dominated.}. In this connection, it is even 
argued in 
refs.~\cite{Brown98,Brown00} that
the effective-to-free-nucleon-mass ratio $m_{\rm N}^*/m_{\rm N}$ is unity
to within a few percent. 
It seems that different model calculations lead to similar conclusions 
about the origin of the Nolen-Schiffer anomaly.

Returning to our model, we note that the results might be improved by
a more self-consistent calculation that takes into account the
feedback(s) between the modified skyrmion and the local nuclear
background and by the inclusion of further degrees of freedom, which
may anyhow be needed from more detailed considerations about the
nucleon structure and the nucleon-nucleon interaction.  
Also the non-local character of the effective nucleon mass may be of
importance. Note that many-body calculations (see
{\em e.g.} \cite{Zu01,Zu05,vD05,vD06,Frick02}) indicate that, while $m^*$ 
is small inside the Fermi sphere, it
can reach values close to the vacuum value near
$k_{\rm F}$.
In this connection, however, we should remark that the gradient
terms which are present in our model do not noticeably affect the
scaling behavior of $m^*$. They are important for the surface behavior
of $\Delta m_{\rm np}^{*}$, though.

\begin{acknowledgement}
We would like to thank Horst Lenske for useful discussions and 
for a computer code calculating 
nuclear densities of finite nuclei 
which allowed us to extract 
the background-density parameters. We are also grateful to 
Frank Gr\"ummer for providing us  with  
calculated shell-model wave functions.
The work of U.T.Y. was supported by the
Alexander von Humboldt Foundation.
The work of A.M.R. was supported by the second phase of
the Brain Korea 21 Project in 2007.
Partial financial support from the EU Integrated Infrastructure
Initiative Hadron Physics Project (contract number RII3-CT-2004-506078),
by the DFG (TR 16, ``Subnuclear Structure of Matter'') and by BMBF
(research grant 06BN411) is gratefully acknowledged.
This work is partially supported by the Helmholtz Association through
funds provided to the virtual institute ``Spin and strong QCD''(VH-VI-231).
\end{acknowledgement}

\begin{appendix}
\section{Mass and moments of inertia of a classical soliton 
in the nonspherical scenario}
\label{app:mass_momin}
\def\theequation{\Alph{section}.\arabic{equation}}
\setcounter{equation}{0}

In the following, we use the  definitions
\begin{eqnarray}
&&P_r\equiv\partial_r P,\quad  P_\theta\equiv\partial_\theta P,\quad
\Theta_r\equiv\partial_r\Theta,\quad\Theta_\theta\equiv\partial_\theta
\Theta,\no\\
&&S_P\equiv \sin P,\quad\!  C_P\equiv \cos P,\quad S_\Theta
\equiv \sin \Theta,\no\\
&&C_\Theta\equiv \cos \Theta,\quad
S_\theta\equiv \sin \theta,\quad\,\, C_\theta\equiv \cos \theta\,.
\end{eqnarray}
The mass of the nonperturbated system  of the Lagrange functional 
 \re{Lag-t} 
has the form
\begin{eqnarray}
M_{\rm NP}^*&=&\pi\intrT 
\times\left\{\dsf{F_\pi^2}{4r^2}\left(1\!-\!\chi_{p}^0(x)\right)\right.\no\\
&\times&\left.
\left[P_\theta^2+r^2P_r^2+S_P^2
\left(\dsf{S_\Theta^2}{S_\theta^2}+\Theta_\theta^2+r^2\Theta_r^2\right)
\right]\right.\no\\
&+&\dsf{S_P^2}{e^2r^4}
\left[\dsf{S_\Theta^2}{S_\theta^2}\left(P_\theta^2+r^2P_r^2\right)\right.\no\\
&+&\left.S_P^2\dsf{S_\Theta^2}{S_\theta^2}
\left(\Theta_\theta^2+r^2\Theta_r^2\right)
+r^2\left(P_r\Theta_\theta-P_\theta\Theta_r\right)^2\right]\no\\
&+&\left.\dsf{m_\pi^2F_\pi^2}{2} \big(1+m_\pi^{-2}\chi_{s}^{00}(x)\big)
(1-C_P)\right\}\,,
\end{eqnarray}
where $x=\sqrt{r^{2}+2r R\cos\theta+R^2}$~\cite{Yakhshiev:2001ht}.
For the moment of inertia, we introduce the following definition 
\begin{eqnarray}
\Lambda&=&2\pi\intrT\,\,\lambda\,, 
\end{eqnarray}
where the contributions of the different parts of the Lagrange 
functional \re{Lag-t} 
are given as
\begin{eqnarray}
\lambda^{*}_{\omega\omega,12}\!&=&\!
\dsf{F_\pi^2}8\left(\!1\!+\!\dsf{\chi_{s}^{02}(x)}{m_\pi^{2}}\right)
\left(1+C_\Theta^2\right) S_P^2\no\\
&+&\!\dsf{S_P^2}{2e^2r^2}
\Big[\left(1+C_\Theta^2\right)\left(P_\theta^2+r^2P_r^2\right)\no\\
&+&S_P^2
\left(\dsf{S_\Theta^2}{S_\theta^2}
+C_\Theta^2\left(\Theta_\theta^2+r^2\Theta_r^2\right)
\right)\Big],\\
\lambda^{*}_{\omega\Omega,12}\!&=&\!\dsf{F_\pi^2}8
\left(\!1\!+\!\dsf{\chi_{s}^{02}(x)}{m_\pi^{2}}\right)\!
\left(C_\theta C_\Theta\dsf{S_\Theta}{S_\theta}
\!+\!\Theta_\theta\!\right)\! S_P^2\no\\
&+&\!\dsf{S_P^2}{2e^2r^2}
\left[C_\theta C_\Theta\dsf{S_\Theta}{S_\theta}
\Big(P_\theta^2\!+\!r^2P_r^2\!+\!
S_P^2\left(\Theta_\theta^2\!+\!r^2P_r^2\right)\!\Big)\right.\no\\
&+&\!\left.r^2P_r(P_r\Theta_\theta-\Theta_rP_\theta)+
S_P^2\dsf{S_\Theta^2}{S_\theta^2}\Theta_\theta\right],\\
\lambda^{*}_{\Omega\Omega,12}\!&=&\!
\dsf{F_\pi^2}8
\left(\!1\!+\!\dsf{\chi_{s}^{02}(x)}{m_\pi^{2}}\right)\!\no\\
&\times&\left[P_\theta^2\!+\!S_P^2
\left(C_\theta^2\dsf{S_\Theta^2}
{S_\theta^2}\!+\!\Theta_\theta^2\right)\!\right]\no\\
&+&\!\dsf{S_P^2}{2e^2r^2}
\left[\dsf{S_\Theta^2}{S_\theta^2}\Big(\left(1+C_\theta^2\right)
\left(P_\theta^2+S_P^2\Theta_\theta^2\right)\right.\no\\
&+&\!\left.r^2\!\left(P_r^2\!+\!S_P^2\Theta_r^2\right)
C_\theta^2\Big)\!+\!r^2\!
\left(P_r\Theta_\theta\!-\!P_\theta\Theta_r\right)^2\right],\\
\lambda^{*}_{\omega\Omega,33}\!&=&\!\dsf{F_\pi^2}4
\left(\!1\!+\!\dsf{\chi_{s}^{02}(x)}{m_\pi^{2}}\right)S_\Theta^2 S_P^2\no\\
&+&\!\dsf{S_P^2}{e^2r^2}
\left(P_\theta^2+r^2P_r^2+S_P^2(\Theta_\theta^2+r^2\Theta_r^2)\right)S_\Theta^2,\\
\lambda_{\rm mes}&=&\dsf{F_\pi^2}{4}S_\Theta^2S_P^2\,,\\
\lambda_{\rm env}^*&=&\dsf{F_\pi^2}{4}\Delta\alpha S_\Theta^2S_P^2\,.\,\,\, 
\end{eqnarray}

\section{Charges and magnetic moments in the  
nonspherical scenario}
\label{app:Charges}

The zero component of the baryonic current 
in nonspherically deformed states has the form
\begin{eqnarray}
B_0&=&-\dsf{S_P^2}{2\pi^2r^2}\,\dsf{\sin\Theta}{\sin\theta}\,
(P_r\Theta_\theta-P_\theta\Theta_r)\,.
\end{eqnarray}
In calculating the third component of the isospin current $V_0^{(3)}$, 
one finds
\begin{eqnarray}
\int\!{\rm d}^3r\,V_0^{(3)}
=(a^*\!+\!\omega_2\!-\!\Omega_3)\Lambda_{\omega\Omega,33}^*\!+\!
\Lambda_{\rm env}^*\!\equiv\! T_3\,.
\end{eqnarray}
Consequently, the angle-averaged isospin charge {\em density} is given as 
\begin{eqnarray}
{\tilde T_3}
&=&
(a^*+\omega_2-\Omega_3)\tilde\lambda_{\omega\Omega,33}^*
+\tilde\lambda_{\rm env}^*\no\\
&=&\dsf{T_3-\Lambda_{\rm env}^*}{\Lambda_{\omega\Omega,33}^*}\,
\tilde\lambda_{\omega\Omega,33}^*+\tilde\lambda_{\rm env}^*\,,
\end{eqnarray}
where 
\begin{eqnarray}
\tilde\lambda_{\omega\Omega,33}^*
&=&\dsf12\int\limits_0^\pi\sin\theta\,{\rm d}\theta
\lambda_{\omega\Omega,33}^*\,,\\
\tilde\lambda_{\rm env}^*&=&
\dsf12\int\limits_0^\pi\sin\theta\,{\rm d}\theta\lambda_{\rm env}^*\,.
\end{eqnarray}
Since the charges of the nucleons are defined as 
\begin{eqnarray}
Q^{\left({\rm p}\atop {\rm n}\right)}&=&\dsf{B}{2}
+T_3^{\left({\rm p}\atop {\rm n}\right)}
\equiv\int\limits\!\!\rho^{\rm S}_{\rm E}(r,\theta){\rm d}^3r\pm
\int\limits\!\!\rho^{\rm V}_{\rm E}(r,\theta){\rm d}^3r\,,\no\\
&& \label{charge}
\end{eqnarray}
the isoscalar and the isovector density distributions have here 
the following form
\begin{eqnarray}
\rho^{\rm S}_{\rm E}&=&\dsf{B_0}2-
\dsf{\Lambda_{\rm env}^*}{\Lambda_{\omega\Omega,33}^*}\,
\tilde\lambda_{\omega\Omega,33}^*+\tilde\lambda_{\rm env}^*\,,\\
\rho^{\rm V}_{\rm E}&=&
\dsf{\tilde\lambda_{\omega\Omega,33}^*}{2\Lambda_{\omega\Omega,33}^*}\,.
\end{eqnarray}
Analogous calculations for the magnetic moments lead to
\begin{eqnarray}
\mu^{\left({\rm p}\atop {\rm n}\right)*}&\equiv&
\int\limits\!\!\rho^{\rm S}_{\rm M}(r,\theta){\rm d}^3r\pm
\int\limits\!\!\rho^{\rm V}_{\rm M}(r,\theta){\rm d}^3r\,,
\end{eqnarray}
where
\begin{eqnarray}
\rho^{\rm S}_{\rm M}&=&\dsf{m_{\rm N}
(1+\Lambda^*_{\rm env})}{4\Lambda_{\omega\Omega,33}^*}\, B_0r^2\sin^2\theta\,,\\
\rho^{\rm V}_{\rm M}&=&\dsf{m_{\rm N}}3\Big[\dsf{F_\pi^2}{4}\big(1-\chi_p^0(x)
\big)S_P^2S_\Theta^2+\dsf{S_P^2}{e^2r^2}\no\\
&\times&
\left(P_\theta^2+r^2P_r^2
+S_P^2(\Theta_\theta^2+r^2\Theta_r^2)\right)S_\Theta^2\Big]\,.
\end{eqnarray}

\end{appendix}

\end{document}